\documentclass[prd,amsmath,amssymb,superscriptaddress,preprintnumbers,twocolumn,nofootinbib,10pt]{revtex4-1}
\usepackage{graphicx}
\usepackage{dcolumn}
\usepackage{bm}
\usepackage{amssymb}
\usepackage{latexsym}
\usepackage{booktabs}
\usepackage{amsmath}
\usepackage{multirow}
\usepackage{url}
\usepackage{footnote}
\usepackage{float}
\usepackage{threeparttable}
\usepackage[colorlinks=true, linkcolor=blue, citecolor=blue]{hyperref}
\usepackage[bottom]{footmisc}

\usepackage[normalem]{ulem}
\usepackage{color}
\usepackage{array}
\usepackage{enumerate}
\usepackage{adjustbox}

\usepackage{makecell}
\usepackage{diagbox}
\usepackage{epstopdf}
\usepackage{epsfig}
\usepackage{longtable}
\usepackage{supertabular}
\usepackage{algorithm}
\usepackage{pifont}
\usepackage{algorithmic}
\usepackage{changepage}
\usepackage{setspace}
\begin{document}


\title{Exploring non-cold dark matter in a scenario of dynamical dark energy with DESI DR2 data}

\author{Tian-Nuo Li}
\affiliation{Liaoning Key Laboratory of Cosmology and Astrophysics, College of Sciences, Northeastern University, Shenyang 110819, China}

\author{Peng-Ju Wu}
\affiliation{School of Physics, Ningxia University, Yinchuan 750021, China}

\author{Guo-Hong Du}
\affiliation{Liaoning Key Laboratory of Cosmology and Astrophysics, College of Sciences, Northeastern University, Shenyang 110819, China}

\author{Yan-Hong Yao}
\affiliation{Institute of Fundamental Physics and Quantum Technology, Department of Physics, School of Physical Science and Technology, Ningbo University, Ningbo, Zhejiang 315211, China}

\author{Jing-Fei Zhang}\thanks{Corresponding author}\email{ jfzhang@mail.neu.edu.cn}
\affiliation{Liaoning Key Laboratory of Cosmology and Astrophysics, College of Sciences, Northeastern University, Shenyang 110819, China}

\author{Xin Zhang}\thanks{Corresponding author}\email{zhangxin@mail.neu.edu.cn}
\affiliation{Liaoning Key Laboratory of Cosmology and Astrophysics, College of Sciences, Northeastern University, Shenyang 110819, China}
\affiliation{MOE Key Laboratory of Data Analytics and Optimization for Smart Industry, Northeastern University, Shenyang 110819, China}
\affiliation{National Frontiers Science Center for Industrial Intelligence and Systems Optimization, Northeastern University, Shenyang 110819, China}

\begin{abstract}

Recent observations of DESI hint that dark matter (DM) may not be cold but have a non‑zero equation of state (EoS) parameter, and that dark energy (DE) may not be the cosmological constant. In this work, we explore the possibility of a non-zero DM EoS parameter within the framework of dynamical DE. We perform analysis by using the latest baryon acoustic oscillation (BAO) data from DESI DR2, the cosmic microwave background (CMB) data from Planck, and the type Ia supernova (SN) data from DESY5 and PantheonPlus. When using the combination of CMB, BAO, and SN data, our results indicate a preference for a non‑zero DM EoS parameter at the $2.8\sigma$ and $3.3\sigma$ level within the content of a constant DE EoS. In contrast, for a time‑evolving DE EoS parameterized by $w_0$ and $w_a$, this preference decreases to $0.8\sigma$ and $1.1\sigma$. Furthermore, allowing a non‑zero DM EoS yields best‑fit values of $w_0$ and $w_a$ that exhibit smaller deviations from the $\Lambda$CDM expectations, and Bayesian evidence analysis shows a comparable preference for this model relative to $\Lambda$CDM. The overall results of this work indicate that a non‑zero DM EoS parameter warrants further exploration and investigation.

\end{abstract}
\maketitle

\section{Introduction}

Observational evidence suggest that the late-time universe is dominated by two major components: dark matter (DM) and dark energy (DE), which together contribute approximately 95\% of the total energy budget of the universe \cite{Planck:2018vyg}. DM drives the formation of cosmic structures \cite{Wechsler:2018pic}, and DE is responsible for the accelerated expansion of the universe \cite{SupernovaSearchTeam:1998fmf,SupernovaCosmologyProject:1998vns}, while the fundamental nature and origin of both components are still unknown. In the standard $\Lambda$CDM model, DE is modeled as a cosmological constant $\Lambda$ with an equation of state (EoS) $w = -1$, while DM is regarded as a cold, pressureless, non-interacting (except gravitationally) perfect fluid with a zero EoS parameter, $w_\mathrm{dm} = 0$. After nearly two decades, the $\Lambda$CDM model provides a good fit to the majority of cosmological observations.

Despite the $\Lambda$CDM model has been successful, it still encounters several issues related to the cosmological constant and cold dark matter (CDM). For the CDM, the observed properties of halos deviate from the predictions made by the CDM framework \cite{Bullock:2017xww}. In order to alleviate some issues of CDM, various DM candidates have been proposed, such as warm DM \cite{Blumenthal:1982mv,Bode:2000gq}, fuzzy DM \cite{Hu:2000ke}, condensate DM \cite{Schive:2014dra}, and decaying DM \cite{Wang:2014ina} (see Ref.~\cite{Bertone:2004pz} for a review). Most DM candidates can be effectively modeled within the Generalized Dark Matter framework, which employs various parameterizations of the EoS parameter, sound speed, and viscosity, as first introduced in Ref.~\cite{hu1998structure}. This approach has motivated several researchers to investigate whether observational data support non-cold DM (i.e., a non-zero EoS parameter of DM) \cite{Muller:2004yb,Kumar:2012gr,Kopp:2016mhm,Murgia:2017lwo,Gariazzo:2017pzb,Murgia:2018now,Kopp:2018zxp,Schneider:2018xba,Kumar:2019gfl,Ilic:2020onu,Najera:2020smt,Pan:2022qrr,Yao:2023ybs,Yao:2025kuz}. For example, to test the warmness of DM, \citet{Muller:2004yb} has investigated the DM EoS parameter using cosmic microwave background (CMB), type Ia supernova (SN), and large scale structure data with zero adiabatic sound speed and no entropy production. Ref.~\cite{Pan:2022qrr} explores the non-zero DM EoS parameter using CMB, baryon acoustic oscillations (BAO), and SN data in the framework of DE and DM interaction, providing evidence at approximately the $1\sigma$ level. Therefore, it is both important and worthwhile in modern cosmology to examine whether the observational data support a deviation from the CDM paradigm, thereby further exploring the nature of DM.

Recently, the measurements of BAO from the second data release (DR2) of the Dark Energy Spectroscopic Instrument (DESI), based on three years of observations, have been published. The combination of DESI DR2 BAO data, CMB data, and SN data reveals a $2.8\sigma-4.2\sigma$ preference for dynamical DE within the $w_0w_a$CDM model \cite{DESI:2025zgx}. These significant deviations from the cosmological constant, as highlighted by DESI, have sparked extensive debates on the nature of DE \cite{Li:2024qso,Giare:2024gpk,Dinda:2024ktd,Escamilla:2024ahl,Sabogal:2024yha,Li:2024qus,Li:2024hrv,Wang:2024dka,Huang:2025som,Li:2025owk,Wu:2025wyk,Li:2025ula,Li:2025ops,Barua:2025ypw,Yashiki:2025loj,Ling:2025lmw,Goswami:2025uih,Yang:2025boq,Pang:2025lvh,You:2025uon,Ozulker:2025ehg,Cheng:2025lod,Pan:2025qwy}. Furthermore, this has also stimulated research considered to constrain various aspects of cosmological physics \cite{Du:2024pai,Jiang:2024viw,Wu:2024faw,Ye:2024ywg,Du:2025iow,Feng:2025mlo,Pan:2025psn,Wang:2025dtk,Cai:2025mas,Li:2025cxn,Odintsov:2025jfq,Nojiri:2025low}, particularly exploring the nature of DM as an important issue \cite{Yang:2025ume, Chen:2025wwn, Wang:2025zri, Kumar:2025etf, Abedin:2025dis, Khoury:2025txd, Araya:2025rqz}. For instance, \citet{Kumar:2025etf} investigated the potential deviations from CDM and found that the current observational data favor a non-zero DM EoS parameter at approximately the $2\sigma$ level. In our recent work \cite{Li:2025eqh}, we explored the non-zero DM EoS within the PEDE framework, which shows a preference at approximately the $3\sigma$ level. These different DE models may influence the investigation of the DM EoS parameter. 

In this work, we explore the possibility of a non-zero DM EoS parameter using BAO from DESI DR2, CMB from Planck, and SN from DESY5 and PantheonPlus. Considering that DE may not be the cosmological constant, we conduct a comprehensive analysis with different parameteration of DE EoS, discussing the impact of DE on the exploration of non-zero DM EoS parameter. Furthermore, we aim to examine the impact of a non-cold DM on the measurement of DE EoS.

This work is organized as follows. In Sec.~\ref{sec2}, we briefly introduce the models considered in this work, along with the cosmological data utilized in the analysis. In Sec.~\ref{sec3}, we report the constraint results and make some relevant discussions. The conclusion is given in Sec.~\ref{sec4}.

\section{methodology and data}\label{sec2}

\subsection{models}\label{sec2.1}

We now present the cosmological models that are investigated in this work. We consider spatially flat, homogeneous, and isotropic cosmologies governed by general relativity, described by the Friedmann–Robertson–Walker (FRW) metric. We assume that the matter components minimally couple to gravity and there are no non-gravitational interactions between radiation, baryons, non-cold DM, and DE. 

In the background level, the dimensionless Hubble parameter is given by

\begin{equation}
\frac{H^2(a)}{H_0^2} = \Omega_{\rm r0}a^{-4} + \Omega_{\rm dm0}a^{-3(1+w_{\rm dm})} + \Omega_{\rm b0}a^{-3} + \Omega_{\rm de0} f(a),
\end{equation}
where $a=1/(1+z)$ is the scale factor, $H(a)$ is the Hubble parameter, $\Omega_{\rm r0}$, $\Omega_{\rm dm0}$, $\Omega_{\rm b0}$, and $\Omega_{\rm de0}$ are the present density parameters for radiation, non-cold DM, baryons, and DE, respectively. $f(a)$ represents the normalized $a$-dependent density of DE, given by
\begin{equation}
f(a) = \exp\left( -3 \int_{1}^{a} \frac{1 + w(a')}{a'} {\rm d}a' \right),
\end{equation}
where $w(a)$ is the EoS of DE.

The main focus of our work is to explore the possibility of a non-zero DM EoS parameter. To test whether this parameter deviates from zero, we consider a free DM EoS (labeled as \textbf{nCDM}) in three DE models: (i) \textbf{$\boldsymbol{\Lambda}$nCDM} model with $w=-1$; (ii) \textbf{$\boldsymbol{w}$nCDM} model with a constant $w$; (iii) \textbf{$\boldsymbol{w_0w_a}$nCDM} model with the Chevallier–Polarski–Linder parameterization form, $w(a)=w_0+w_a(1-a)$.

In the linear perturbation level, the perturbed FRW metric with the conformal Newtonian gauge takes the form
\begin{equation}
  {\rm d}s^2 = a^2(\tau) \left[ -(1+2\psi) {\rm d}\tau^2 + (1-2\phi) {\rm d}\vec{r}^2 \right],
\end{equation}
where $\tau$ is conformal time, $\psi$ and $\phi$ are the Bardeen potentials ($\phi = \psi$ in the absence of anisotropic stress), and $\vec{r}$ is the comoving spatial coordinate. The linear evolution of density and velocity perturbations for DM and DE is governed by the conservation equations derived from the first-order perturbations of the stress-energy-momentum tensor. In Fourier space, these are given by the continuity and Euler equations

\begin{widetext}
\begin{equation}
 \delta^{\prime}_{\rm ds}= -(1+w_{\rm ds}) \left(\theta_{\rm ds} - 3 \phi^{\prime} \right)
 -3 \mathcal{H} \delta_{\rm ds} (c^2_{\rm s,ds} - w_{\rm ds}) - 9 (1+w_{\rm ds})(c^2_{\rm s,ds} - c^2_{\rm a,ds})\mathcal{H}^2 \frac{\theta_{\rm ds}}{k^2},
\end{equation}
\begin{equation}
\theta^{\prime}_{\rm ds}=-(1-3 c^2_{\rm s,ds}) \mathcal{H} \theta_{\rm ds}  + \frac{c^2_{\rm s,ds}}{1+w_{\rm ds}}k^2 \delta_{\rm ds} + k^2\psi.
\end{equation}

\end{widetext}
Here, subscript ``ds'' represents the dark sector, including DE and DM, thus ds = de or dm. The prime denotes the derivative with respect to conformal time, $\mathcal{H} = a^{\prime}/a$ is the conformal Hubble parameter, and $k$ is the magnitude of the wavevector. $\delta_{\rm ds}$ and $\theta_{\rm ds}$ represent the density perturbations and the velocity divergence perturbations, respectively. $c_{\rm s,ds}$ and $c_{\rm a,ds}$ are the sound speed and adiabatic sound speed for dark sector. For the sake of simplicity, we fix $c^2_{\rm s,de} = 1$ and the sound speed of non-cold DM is set to  $c^2_{\rm s,dm} = 0$.

\subsection{Cosmological data}\label{sec2.2}

\begin{table}[t]
\caption{Flat priors on the main cosmological parameters constrained in this paper.}
\begin{center}
\renewcommand{\arraystretch}{1.1}
\begin{tabular}{@{\hspace{0.8cm}} c @{\hspace{1.0cm}} c @{\hspace{0.8cm}}}
\hline\hline
 \textbf{Parameter}       & \textbf{Prior}\\
\hline
$\omega_{\rm b}$                  & $\mathcal{U}$[0.005\,,\,0.1] \\
$\omega_{\rm dm}$                 & $\mathcal{U}$[0.01\,,\,0.99] \\
$\theta_{\rm s}$                  & $\mathcal{U}$[0.5\,,\,10] \\
$\tau_{\rm reio}$                 & $\mathcal{U}$[0.01\,,\,0.8] \\
$\log(10^{10}A_{\rm s})$          & $\mathcal{U}$[1.61\,,\,3.91] \\
$n_{\rm s}$                       & $\mathcal{U}$[0.8\,,\,1.2] \\
$w~{\rm or}~w_0$                  & $\mathcal{U}$[-3\,,\,1] \\
$w_a$                             & $\mathcal{U}$[-3\,,\,2] \\
$w_\mathrm{dm}$                   & $\mathcal{U}$[-0.1\,,\,0.1] \\
\hline\hline
\end{tabular}
\label{tab1}
\end{center}	
\end{table}

We list the free parameters of these models and the uniform priors applied in Table~\ref{tab1}. The parameter set for the $\Lambda$CDM model is $\bm{\theta}_{\Lambda\mathrm{CDM}}=\{\omega_{\rm b}$, $\omega_{\rm dm}$, $\log(10^{10} A_{\mathrm{s}})$, $\theta_{\rm s}$, $n_{\mathrm{s}}$, $\tau_{\rm reio}\}$. For the extended models, the parameter sets are $\bm{\theta}_{\Lambda\mathrm{nCDM}}=\{\bm{\theta}_{\Lambda\mathrm{CDM}},w_{\rm dm}\}$, $\bm{\theta}_{w\mathrm{nCDM}}=\{\bm{\theta}_{\Lambda\mathrm{CDM}},w_{\rm dm},w\}$, and $\bm{\theta}_{w_0w_a\mathrm{nCDM}}=\{\bm{\theta}_{\Lambda\mathrm{CDM}},w_{\rm dm},w_0,w_a\}$. We compute the theoretical model using a modified version of the {\tt CLASS} code \cite{Lesgourgues:2011re,Blas:2011rf}. We perform Markov Chain Monte Carlo (MCMC) \cite{Lewis:2002ah,Lewis:2013hha} 
analysis using the publicly available sampler {\tt Cobaya}\footnote{\url{https://github.com/CobayaSampler/cobaya}} \cite{Torrado:2020dgo} and assess the convergence of the MCMC chains using the Gelman-Rubin statistics quantity $R - 1 < 0.02$ \cite{Gelman:1992zz}. The MCMC chains are analyzed using the public package {\tt GetDist}\footnote{\url{https://github.com/cmbant/getdist/}} \cite{Lewis:2019xzd}. We use the current observational data to constrain these models and obtain the best-fit values and the $1$--$2\sigma$ confidence level ranges for the parameters of interest \{$H_{0}$, $\Omega_{\mathrm{m}}$, $w_{\rm dm}$, $w$ or $w_0$, $w_a$\}. 

The datasets used are as follows:
\begin{itemize}
    \item \textbf{\texttt{CMB}:} The CMB likelihoods consist of four components: (i) the small-scale ($\ell>30$) temperature and polarization power spectra, $C_\ell^{TT}$, $C_\ell^{TE}$ and $C_\ell^{EE}$, obtained from the Planck \texttt{CamSpec} likelihood~\cite{Planck:2018vyg,Efstathiou:2019mdh,Rosenberg:2022sdy}; (ii) the large-scale ($2 \le \ell \le 30$) temperature spectrum, $C_\ell^{TT}$, from the Planck \texttt{Commander} likelihood~\cite{Planck:2018vyg,Planck:2019nip}; (iii) the large-scale ($2 \le \ell \le 30$) E-mode polarization spectrum, $C_\ell^{EE}$, from the Planck \texttt{SimAll} likelihood~\cite{Planck:2018vyg,Planck:2019nip}; (iv) the CMB lensing likelihood, utilizing the latest high-precision reconstruction from NPIPE PR4 Planck data\footnote{Available at \url{https://github.com/carronj/planck_PR4_lensing}.}~\cite{Carron:2022eyg}.

    \item \textbf{\texttt{DESI}:} The BAO measurements from DESI DR2 include tracers of the bright galaxy sample, luminous red galaxies, emission line galaxies, quasars, and the Lyman-$\alpha$ forest. These measurements are described through the transverse comoving distance $D_{\mathrm{M}}/r_{\mathrm{d}}$, the angle-averaged distance $D_{\mathrm{V}}/r_{\mathrm{d}}$, and the Hubble horizon $D_{\mathrm{H}}/r_{\mathrm{d}}$, where  $r_{\mathrm{d}}$ is the comoving sound horizon at the drag epoch. The measurements we use detailed in Table IV of Ref.~\cite{DESI:2025zgx}.

    \item \textbf{\texttt{PantheonPlus}:} The PantheonPlus comprises 1550 spectroscopically confirmed type Ia supernovae (SNe) from 18 different surveys, covering $0.01 < z < 2.26$ \cite{Brout:2022vxf}\footnote{Data available at \url{https://github.com/PantheonPlusSH0ES/DataRelease}}.

    \item \textbf{\texttt{DESY5}:}  The DESY5 sample comprises 1635 photometrically classified SNe from the released part of the full 5 yr data of the Dark Energy Survey collaboration (with redshifts in the range $0.1 < z < 1.3$), complemented by 194 low-redshift SNe from the CfA3~\cite{Hicken:2009df}, CfA4~\cite{Hicken:2012zr}, CSP~\cite{Krisciunas:2017yoe}, and Foundation~\cite{Foley:2017zdq} samples (with redshifts in the range $0.025 < z < 0.1$), for a total of 1829 SNe \cite{DES:2024jxu}.\footnote{Data available at \url{https://github.com/des-science/DES-SN5YR}.}

\end{itemize}

\section{Results and discussions}\label{sec3}

\begin{table*}[!htb]
\centering
\caption{Fitting results ($1\sigma$ confidence level) in the $\Lambda$CDM, $\Lambda$nCDM, $w$CDM, $w$nCDM, $w_0w_a$CDM, and $w_0w_a$nCDM models from the CMB+DESI, CMB+DESI+DESY5, and CMB+DESI+Pantheonplus data. Here, $H_{0}$ is in units of ${\rm km}~{\rm s}^{-1}~{\rm Mpc}^{-1}$.}
\label{tab2}
\setlength{\tabcolsep}{2mm}
\renewcommand{\arraystretch}{1.2}
\footnotesize
\begin{tabular}{lc c c c c c c}
\hline 
\hline
Model/Dataset & $H_0$ &$\Omega_{\mathrm{m}}$& $w_\mathrm{dm}$ & $w$ or $w_0$ & $w_a$ \\
\hline
$\bm{\Lambda}$\textbf{CDM} &  &  &  &  &  \\
CMB+DESI & $68.19^{+0.30}_{-0.27}$ & $0.3025\pm 0.0037$ & --- & --- & --- \\
CMB+DESI+DESY5 & $68.03\pm 0.27$ & $0.3047\pm 0.0035$ & --- & --- & --- \\
CMB+DESI+PantheonPlus & $68.12\pm 0.27$ & $0.3035\pm 0.0035$ & --- & --- & --- \\
\hline
$\bm{\Lambda\textbf{nCDM}}$ &  &  &  &  &  \\
CMB+DESI & $68.93\pm 0.42$ & $0.2949\pm 0.0047$  & $0.00084\pm 0.00035$ & --- & --- \\
CMB+DESI+DESY5 & $68.52\pm 0.40$ & $0.2996\pm 0.0046$  & $0.00059\pm 0.00033$ & --- & --- \\
CMB+DESI+PantheonPlus & $68.73\pm 0.40$ & $0.2971\pm 0.0046$  & $0.00074\pm 0.00034$ & --- & --- \\
\hline
$\bm{w}$\textbf{CDM} &  &  &  &  &  \\
CMB+DESI & $69.42^{+0.86}_{-0.98}$ & $0.2933\pm 0.0073$ & --- & $-1.050^{+0.039}_{-0.033}$ & --- \\
CMB+DESI+DESY5 & $67.33\pm 0.52$ & $0.3097\pm 0.0048$ & --- & $-0.969\pm 0.020$ & --- \\
CMB+DESI+PantheonPlus & $67.95\pm 0.55$ & $0.3048\pm 0.0049$ & --- & $-0.993\pm 0.022$ & --- \\
\hline
$\bm{{w}\textbf{nCDM}}$ &  &  &  &  &  \\
CMB+DESI & $68.54^{+0.91}_{-1.03}$ & $0.2972\pm 0.0072$ & $0.00103\pm 0.00051$ & $-0.978^{+0.051}_{-0.045}$ & --- \\
CMB+DESI+DESY5 & $67.30\pm 0.53$ & $0.3056^{+0.0046}_{-0.0051}$ & $0.00147\pm 0.00045$ & $-0.916\pm 0.026$ & --- \\
CMB+DESI+PantheonPlus & $67.89\pm 0.58$ & $0.3016\pm 0.0052$ & $0.00124\pm 0.00044$ & $-0.946\pm 0.027$ & --- \\

\hline
$\bm{w_0w_a}$\textbf{CDM} &  &  &  &  &  \\
CMB+DESI & $63.50^{+1.51}_{-2.02}$ & $0.3540^{+0.0210}_{-0.0180}$ & --- & $-0.410^{+0.210}_{-0.180}$ & $-1.77\pm 0.54$ \\
CMB+DESI+DESY5 & $66.72\pm 0.54$ & $0.3189\pm 0.0055$ & --- & $-0.756\pm 0.059$ & $-0.83^{+0.24}_{-0.21}$ \\
CMB+DESI+PantheonPlus & $67.54^{+0.56}_{-0.63}$ & $0.3109\pm 0.0057$ & --- & $-0.844\pm 0.054$ & $-0.58\pm 0.20$ \\
\hline
$\bm{{w_0w_a}\textbf{nCDM}}$ &  &  &  &  &  \\
CMB+DESI & $63.80\pm 2.03$ & $0.3510^{+0.0230}_{-0.0260}$ & $0.00005^{+0.00044}_{-0.00064}$ & $-0.450\pm 0.230$ & $-1.63\pm 0.71$ \\
CMB+DESI+DESY5 & $66.88\pm 0.58$ & $0.3157^{+0.0060}_{-0.0069}$ & $0.00040^{+0.00050}_{-0.00057}$ & $-0.781\pm 0.061$ & $-0.68\pm 0.28$ \\
CMB+DESI+PantheonPlus & $67.63^{+0.61}_{-0.51}$ & $0.3077^{+0.0055}_{-0.0067}$ & $0.00058^{+0.00059}_{-0.00052}$ & $-0.862^{+0.056}_{-0.064}$ & $-0.43^{+0.32}_{-0.23}$ \\
\hline
\hline
\end{tabular}
\end{table*}

\begin{figure}[H]
\includegraphics[scale=0.45]{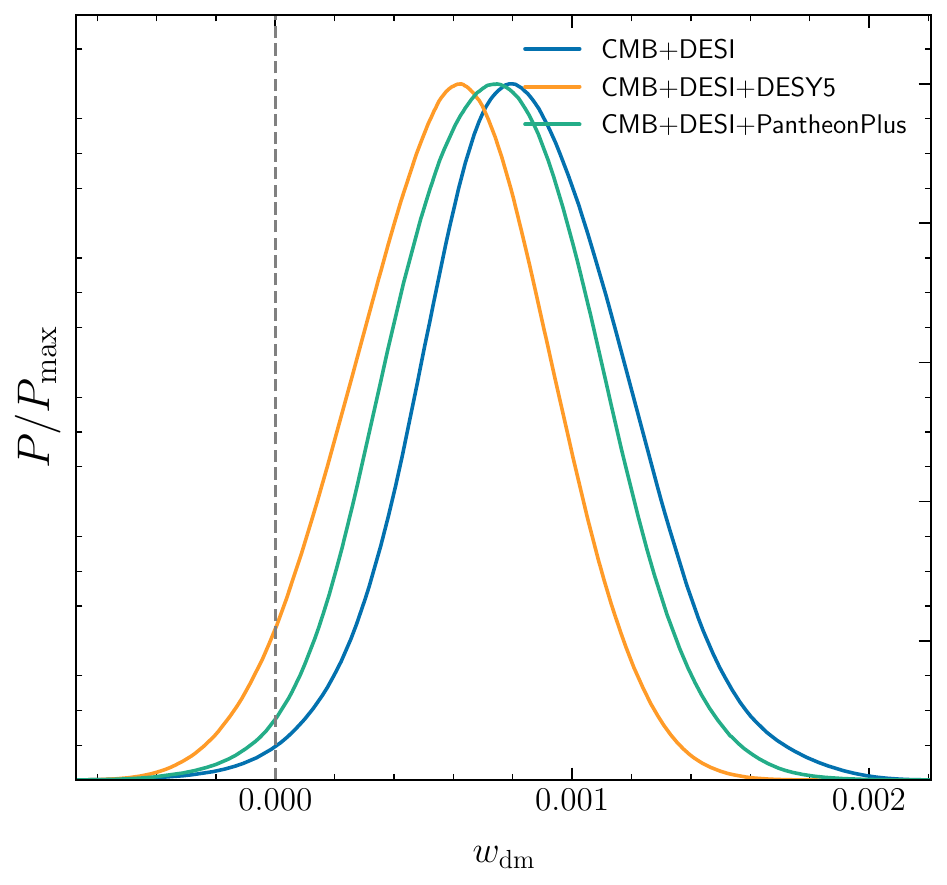}
\centering
\caption{\label{fig1} The 1D marginalized posterior constraints on $w_{\rm dm}$ from CMB+DESI, CMB+DESI+DESY5, and CMB+DESI+pantheonPlus data in the $\Lambda$nCDM model.}
\end{figure}

\begin{figure*}[htbp]
\centering
\includegraphics[width=0.44\textwidth, height=0.31\textheight]{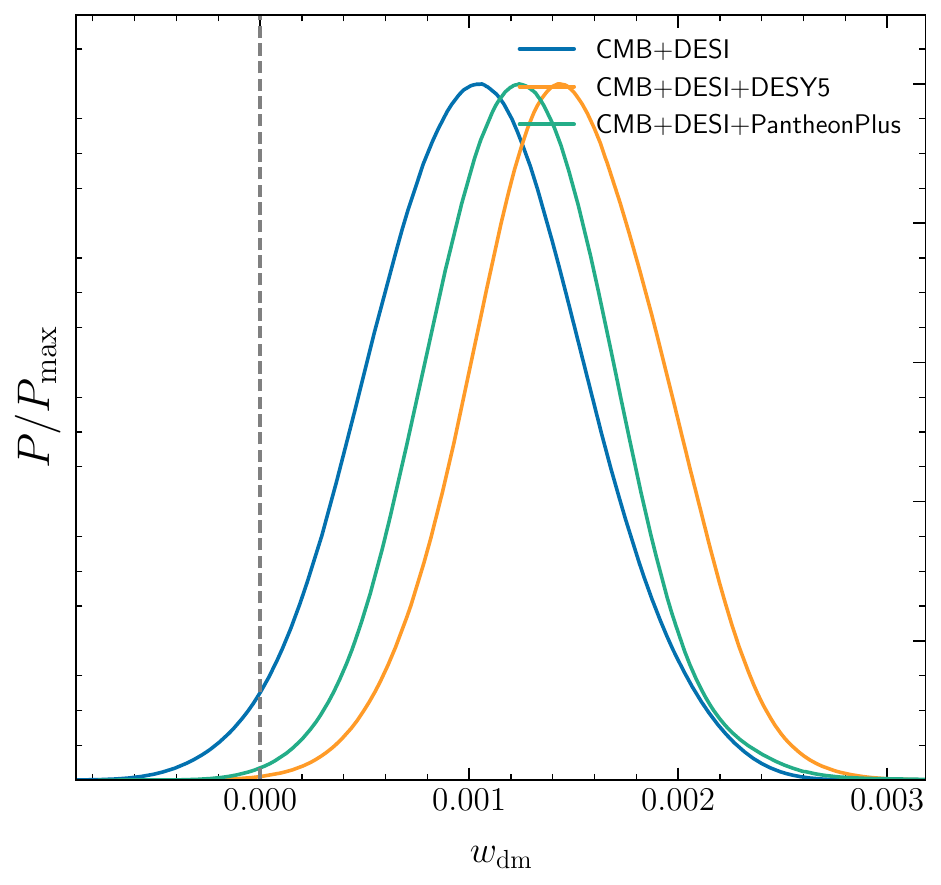} \hspace{1cm}
\includegraphics[width=0.46\textwidth, height=0.315\textheight]{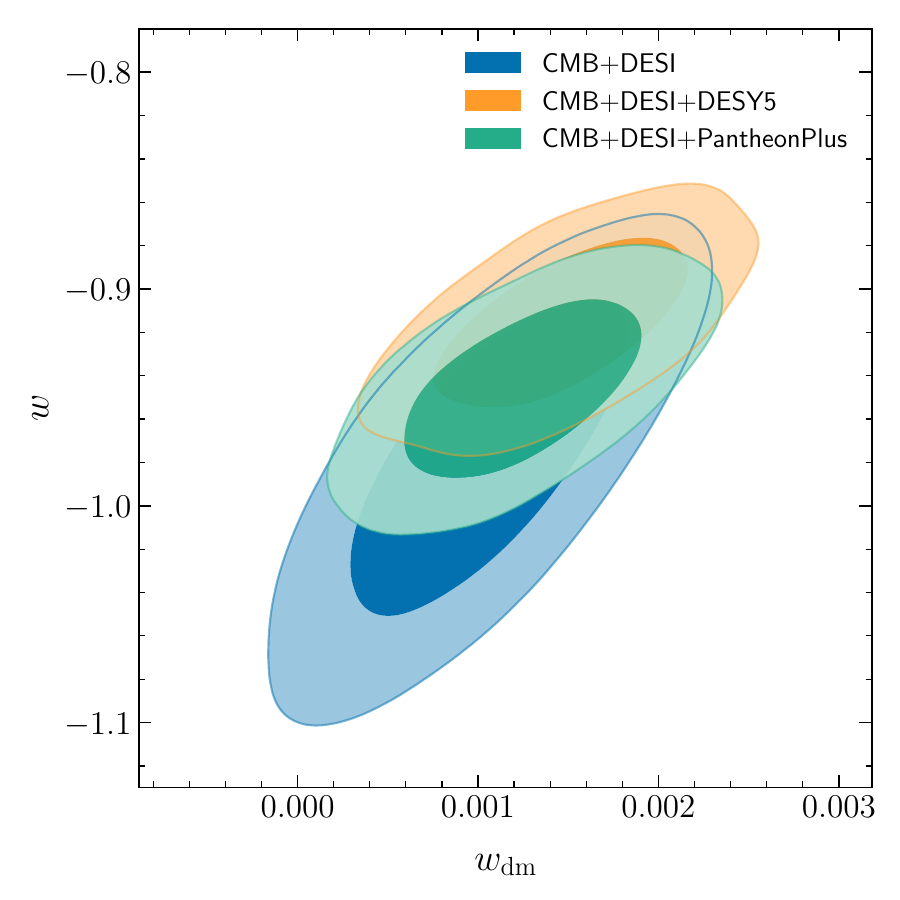}
\caption{\label{fig2} Constraints on the cosmological parameters using the CMB, DESI, DESY5, and PantheonPlus data in the $w$nCDM model. \emph{Left panel}: The marginalized 1D posteriors on $w_{\rm dm}$ in $w$nCDM model, from CMB+DESI, CMB+DESI+DESY5, and CMB+DESI+PantheonPlus as labelled. \emph{Right panel}: Two-dimensional marginalized contours ($1\sigma$ and $2\sigma$ confidence levels) in the $w_{\rm dm}$--$w$ plane by using the CMB+DESI, CMB+DESI+DESY5, and CMB+DESI+PantheonPlus data in the $w$nCDM model.}
\end{figure*}

\begin{figure}[htbp]
\includegraphics[scale=0.5]{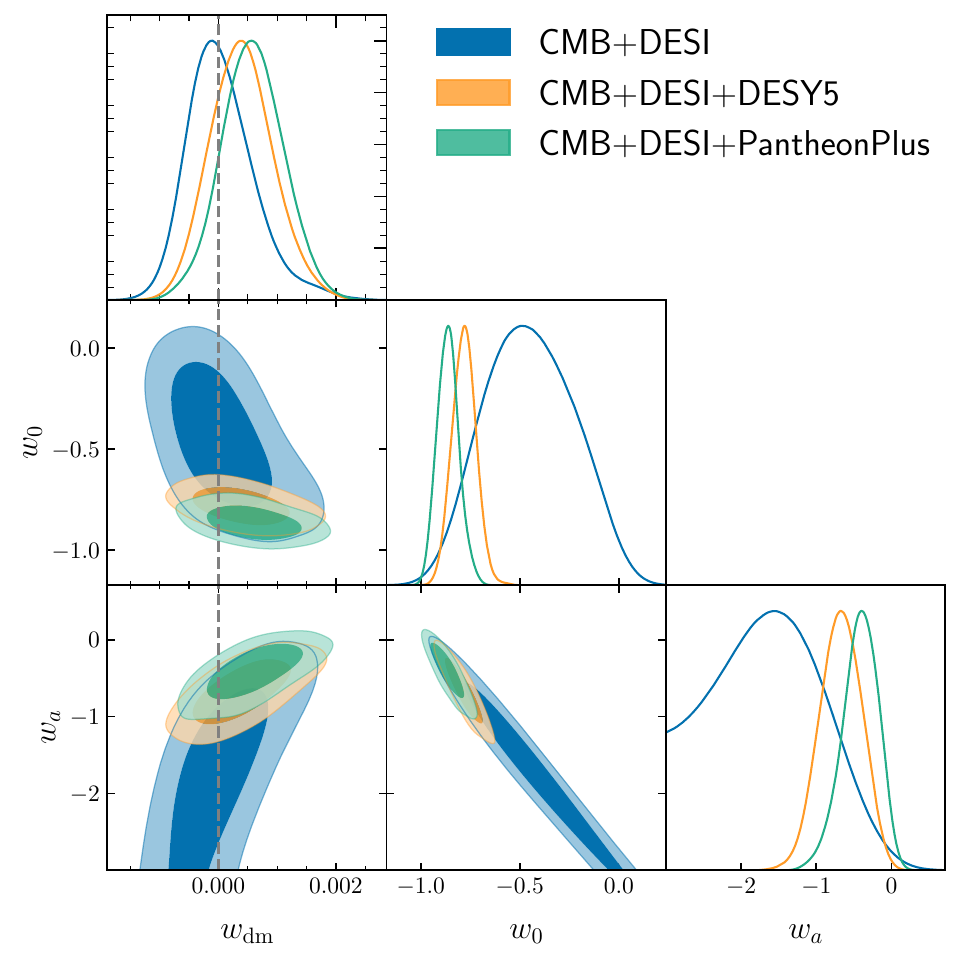}
\centering
\caption{\label{fig3} Constraints on the cosmological parameters using the CMB+DESI, CMB+DESI+DESY5, and CMB+DESI+PantheonPlus data in the $w_0w_a$nCDM model.}
\end{figure}

\begin{figure*}[htbp]
\includegraphics[scale=0.45]{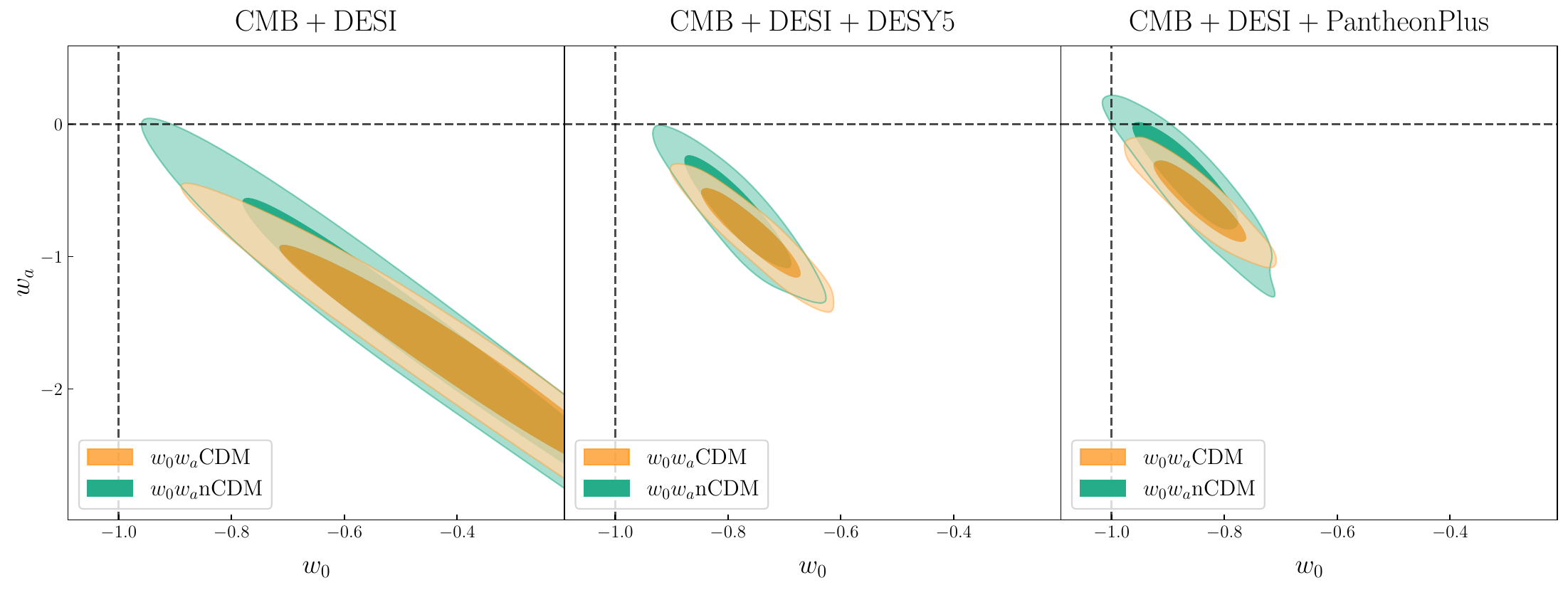}
\centering
\caption{\label{fig4} Comparison of the constraints on $w_0$ and $w_a$ in the $w_0w_a$CDM and $w_0w_a$nCDM models using the CMB+DESI, CMB+DESI+DESY5, and CMB+DESI+PantheonPlus data.}
\end{figure*}

\begin{figure}[htbp]
\includegraphics[scale=0.35]{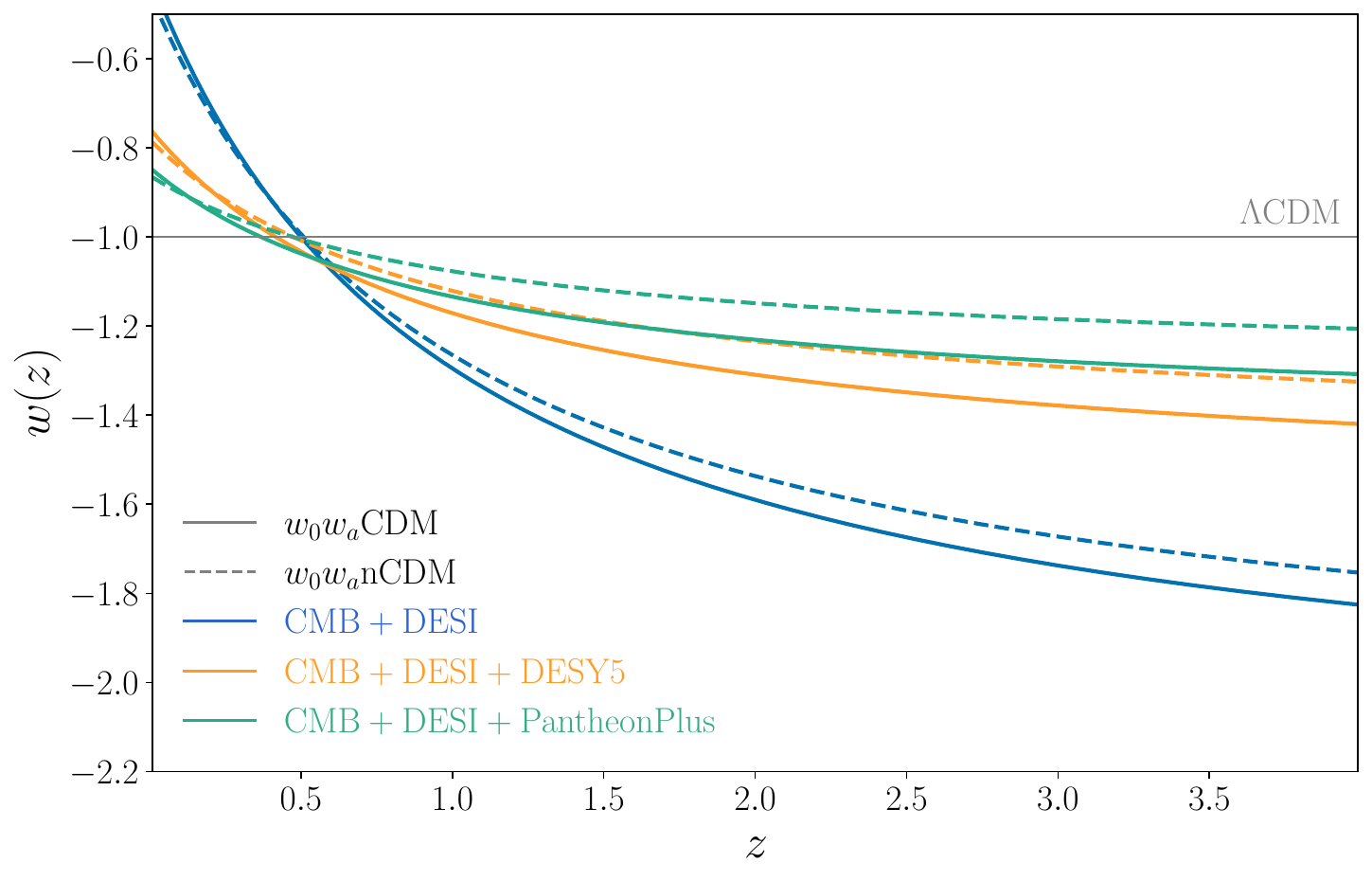}
\centering
\caption{\label{fig5} The evolution of DE EoS $w(z)$ for $w_0w_a$CDM and $w_0w_a$nCDM models from the CMB+DESI, CMB+DESI+DESY5, and CMB+DESI+Pantheonplus data.}
\end{figure}

In this section, we shall report the constraint results of the cosmological parameters. We consider the $\Lambda$CDM, $\Lambda$nCDM, $w$CDM, $w$nCDM, $w_0w_a$CDM, and $w_0w_a$nCDM models to perform a cosmological analysis using current observational data, including the DESI, CMB, DESY5, and PantheonPlus data. We show the $1\sigma$ and $2\sigma$ posterior distribution contours for various cosmological parameters in the $\Lambda$nCDM, $w$nCDM, and $w_0w_a$nCDM models in Figs.~\ref{fig1}--\ref{fig3}. The $1\sigma$ errors for the marginalized parameter constraints are summarized in Table~\ref{tab2}. We compare the two-dimensional marginalized contours $w_0$--$w_a$ and the EoS of DE for the $w_0w_a$CDM and $w_0w_a$nCDM models using the DESI, CMB, DESY5, and PantheonPlus datasets, as shown in Figs.~\ref{fig4} and \ref{fig5}. Finally, we list the Bayes factors of those models in Table~\ref{tab3}.

In Fig.~\ref{fig1}, we present the constraint results on the one-dimensional (1D) marginalized posterior distributions of $w_{\rm dm}$ for the $\Lambda$nCDM model, using the current observational data. The constraint values of $w_{\rm dm}$ are $0.00084\pm 0.00035$ (CMB+DESI), $0.00059\pm 0.00033$ (CMB+DESI+DESY5), and $0.00074\pm 0.00034$ (CMB+DESI+PantheonPlus), indicating a preference for a non-zero DM EoS parameter at $2.4\sigma$, $1.8\sigma$, and $2.2\sigma$ confidence levels, respectively. Note that constraints on $\Lambda$nCDM have been presented in Ref.~\cite{Kumar:2025etf} using current cosmological data, and these results are largely consistent with ours. However, due to the differences in the CMB data and the inclusion of the PR4 lensing data in our analysis, the constraint results are slightly improved, and a slightly higher deviation is also observed.

Next, we investigate how the parameter $w$ affects the inference of the DM EoS parameter ($w_{\rm dm}$) in the $w$nCDM model. In the left panel of Fig.~\ref{fig2}, we present the constraint results for the 1D marginalized posterior distributions of $w_{\rm dm}$ in the $w$nCDM model, using the current observational data. The constraint values of $w_{\rm dm}$ are $0.00103 \pm 0.00051$ (CMB+DESI), $0.00147 \pm 0.00045$ (CMB+DESI+DESY5), and $0.00124 \pm 0.00044$ (CMB+DESI+PantheonPlus). These results indicate that the deviation of $w_{\rm dm}$ from zero reaches a significance of $3.3\sigma$ for the CMB+DESI+DESY5 dataset combination, followed by $2.8\sigma$ for CMB+DESI+PantheonPlus, and $2\sigma$ for CMB+DESI. We find that, within the framework of the $w$nCDM model, all data combinations consistently favor slightly higher central values of $w_{\rm dm}$ compared to those inferred from the $\Lambda$nCDM model. This is due to the positive correlation between $w$ and $w_{\rm dm}$ in the $w$nCDM model, and all constraints yield a central value of $w>-1$, as shown in the right panel of Fig.~\ref{fig2}.

We further investigate the impact of exploring a non-zero DM EoS parameter within the framework of a time-evolving DE EoS. In Fig.~\ref{fig3}, we show the triangular plot of the constraint results for the $w_0w_a$nCDM model using the current observational data. Interestingly, CMB+DESI gives $w_{\rm dm} = 0.00005^{+0.00044}_{-0.00064}$, with no evidence for a non-zero DM EoS parameter being found. When the SN data are included, CMB+DESI+DESY5 and CMB+DESI+PantheonPlus give $w_{\rm dm} = 0.00040^{+0.00050}_{-0.00057}$ and $0.00058^{+0.00059}_{-0.00052}$, indicating a preference for a non-zero DM EoS at the $0.8\sigma$ and $1.1\sigma$ levels, respectively. One can easily find that considering an evolving DE EoS leads to the fact that the evidence for a non-zero DM EoS parameter is clearly weakened by $1\sim2\sigma$ compared to that in the $\Lambda$nCDM and $w$nCDM models.

Additionally, we have also explored the impact of considering a free DM EoS parameter, $w_{\rm dm}$, on dynamical DE. In Fig.~\ref{fig4}, we present a comparison of the constraint results between the $w_0w_a$CDM model and the $w_0w_a$nCDM model in the $w_0$--$w_a$ plane, based on the current observational data. We find that the contour plots exhibit a slight shift towards the upper left, with the most noticeable effect observed in the CMB+DESI+PantheonPlus combination, followed by CMB+DESI+DESY5, and a very weak shift in CMB+DESI. For example, using the CMB+DESI+PantheonPlus dataset, the $w_{0}w_{a}\mathrm{nCDM}$ model yields $w_{0}=-0.862^{+0.056}_{-0.064}$ and $w_{a}=-0.43^{+0.32}_{-0.23}$, these values are both closer to the $\Lambda$CDM expectations ($w_{0}=-1$, $w_{a}=0$) than those from the $w_{0}w_{a}$CDM model, which yields $w_{0}=-0.844\pm0.054$ and $w_{a}=-0.58\pm0.20$. To more clearly illustrate the impact on DE, the results of the DE evolution with redshift for the $w_0w_a$CDM and $w_0w_a$nCDM models, using CMB+DESI, CMB+DESI+DESY5, and CMB+DESI+PantheonPlus data, are shown in Fig.~\ref{fig5}. For all datasets, we find that the $w_{0}w_{a}\mathrm{nCDM}$ model exhibits a milder evolution of $w(z)$ than the $w_{0}w_{a}$CDM model. Therefore, the $w(z)$ curve crosses the phantom divide (i.e., $w = -1$) earlier than in the $w_0w_a$CDM model.

\begin{table}
\centering
\caption{Summary of the $\ln \mathcal{B}_{ij}$ (where $i$ = $\Lambda$nCDM, $w$nCDM, or $w_0w_a$nCDM; $j$ = $\Lambda$CDM) values quantifying the evidence of these models relative to the $\Lambda$CDM model using current observational datasets. A negative value indicates a preference for the $\Lambda$CDM model.}
\label{tab3}                        
\setlength{\tabcolsep}{2mm}
\renewcommand{\arraystretch}{1.2}
\footnotesize
\begin{tabular}{@{\hspace{0.6cm}}l@{\hspace{0.8cm}} c @{\hspace{0.6cm}}}
\hline 
\hline
Model/Dataset & $\ln \mathcal{B}_{ij}$ \\
\hline

$\bm{\Lambda\textbf{nCDM}}$ &  \\
CMB+DESI & $-2.87$  \\
CMB+DESI+DESY5 & $-4.05$  \\
CMB+DESI+PantheonPlus & $-3.57$ \\
\hline

$\bm{{w}\textbf{nCDM}}$ &  \\
CMB+DESI & $-6.95$  \\
CMB+DESI+DESY5 & $-3.87$  \\
CMB+DESI+PantheonPlus & $-6.42$ \\

\hline

$\bm{{w_0w_a}\textbf{nCDM}}$ &  \\
CMB+DESI & $-4.53$  \\
CMB+DESI+DESY5 & $-0.97$  \\
CMB+DESI+PantheonPlus & $-6.59$ \\
\hline
\hline
\end{tabular}
\end{table}

Finally, we employ the Bayesian Evidence selection criterion as a method for selecting the best model. Here, we use the publicly available code
{\tt MCEvidence}\footnote{\url{https://github.com/yabebalFantaye/MCEvidence}} \cite{Heavens:2017hkr,Heavens:2017afc} to compute the Bayes factor of the models. The Bayes factor is given by $\ln \mathcal{B}_{ij} = \ln Z_i - \ln Z_j$ in logarithmic space, where $Z_i$ and $Z_j$ are Bayesian evidence of two models. Typically, we employ the Jeffreys scale \cite{Kass:1995loi,Trotta:2008qt} to gauge the strength of model preference: $\left|\ln \mathcal{B}_{ij}\right|<1$ (inconclusive evidence); $1\le\left|\ln \mathcal{B}_{ij}\right|<2.5$ (weak evidence); $2.5\le\left|\ln \mathcal{B}_{ij}\right|<5$ (moderate evidence); $5\le\left|\ln \mathcal{B}_{ij}\right|<10$ (strong evidence); and $\left|\ln \mathcal{B}_{ij}\right|\ge 10$ (decisive evidence). 

In Table.~\ref{tab3}, we show the Bayes factors $\ln \mathcal{B}_{ij}$ for the $\Lambda$nCDM, $w$nCDM, and $w_0w_a$nCDM models relative to the $\Lambda$CDM model, based on the current observational data. Here, $i$ denotes the $\Lambda$nCDM, $w$nCDM, or $w_0w_a$nCDM model, and $j$ denotes the $\Lambda$CDM model. It is worth emphasizing that negative values indicate a preference for the $\Lambda$CDM model. For the $w_0w_a$nCDM model, the Bayes factor obtained from CMB+DESI+DESY5 is $\ln \mathcal{B}_{ij} = -0.97$, indicating comparable favor to the $\Lambda$CDM model. However, in all other cases, there is moderate to strong evidence in favor of the $\Lambda$CDM model.

\section{Conclusion}\label{sec4}

In this work, we aim to investigate the potential for a non-zero DM EoS parameter by utilizing the latest BAO data from DESI DR2, CMB data from Planck, and SN data from DESY5 and PantheonPlus. We examine the impact of the DE EoS parameteration on the measurement of DM EoS parameter. Additionally, we intend to evaluate the influence of non‑cold DM on measurements of the DE EoS.

Our analysis reveals a clear preference for a non-zero DM EoS parameter in most of the constraint cases. In the $\Lambda$nCDM model, CMB+DESI shows a $2.4\sigma$ preference for a non-zero DM EoS parameter, which slightly decreases to around $2\sigma$ with the addition of SN data. In the $w$nCDM model, the preference for a non-zero DM EoS parameter significantly increases relative to the $\Lambda$nCDM model, reaching a $3.3\sigma$ level with the CMB+DESI+DESY5 data. The stronger preference arises from the tendency of $w$ to be greater than $-1$, which leads to a more significant deviation of $w_{\rm dm}$ from zero, as $w$ and $w_{\rm dm}$ exhibit a positive correlation. In the $w_0w_a$nCDM model, the evidence for a non-zero DM EoS parameter is clearly weakened by $1 \sim 2\sigma$ compared to the $\Lambda$nCDM and $w$nCDM models. In particular, CMB+DESI yields $w_{\rm dm} = 0.00005^{+0.00044}_{-0.00064}$, showing no preference for a non-zero DM EoS parameter. The overall outcomes clearly indicate that the preference for a non-zero DM EoS parameter is significantly influenced by the DE EoS. 

In addition, we explore the impact of a non-zero DM EoS parameter on the measurement of DE EoS. We find that the $w_0w_a$nCDM model yields values for $w_0$ and $w_a$ that are closer to the $\Lambda$CDM expectations and the $w(z)$ curve crosses the phantom divide at an earlier redshift, compared to the $w_0w_a$CDM model. The Bayesian evidence indicates that the $w_0w_a$nCDM model is comparably favored to the $\Lambda$CDM model using the CMB+DESI+DESY5 data. Based on the results summarized above, the cosmology of a non-zero DM EoS parameter appears more intriguing, and it warrants further exploration with additional high-precision observational data in the future.

\section*{Acknowledgments}
We thank Jing-Zhao Qi, Sheng-Han Zhou, Jia-Le Ling, Yi-Min Zhang, and Ji-Guo Zhang for their helpful discussions. This work was supported by the National SKA Program of China (Grants Nos. 2022SKA0110200 and 2022SKA0110203), the National Natural Science Foundation of China (Grants Nos. 12473001, 11975072, 11875102, and 11835009), the China Manned Space Program (Grant No. CMS-CSST-2025-A02), the National 111 Project (Grant No. B16009), and Guangdong Basic and Applied Basic Research Foundation (Grant No. 2024A1515012573).

\bibliography{main}

\end{document}